\documentclass{aa}
\usepackage{graphics}
\usepackage{txfonts}

\newcommand{\ms}{m\,s$^{-1}$} 
\newcommand{\kms}{km\,s$^{-1}$} 
\newcommand{\te}{$T_{\rm eff}$}

\newcommand{\lgg}{$\log g$}

\newcommand{\vs}{$v_{\rm e}\sin\,i$}

\newcommand{\msun}{$\cal{M}_{\sun}$}

\newcommand{\hg}{\ion{Hg}{ii}}
\newcommand{\hglin}{\ion{Hg}{ii} 3984~\AA}
\newcommand{\aand}{$\alpha$\,And}

\newcommand{\figps}[1]{\resizebox{\hsize}{!}{\rotatebox{0}{\includegraphics{#1}}}}

\begin{document}

\title{Inhomogeneous distribution of mercury on the surfaces of rapidly rotating HgMn stars}

\author{O. Kochukhov\inst{1} \and N. Piskunov\inst{1} \and M. Sachkov\inst{2} \and D. Kudryavtsev\inst{3}}

\offprints{O. Kochukhov \email{oleg@astro.uu.se}}

\institute{Department of Astronomy and Space Physics, Uppsala University, SE-751 20, Uppsala, Sweden
      \and Institute of Astronomy, Russian Academy of Sciences, Pyatnitskaya 48, 119017 Moscow, Russia
      \and Special Astrophysical Observatory, Russian Academy of Sciences, Nizhnii Arkhyz, 357147 
           Karachai-Cherkessian Republic, Russia}

\date{Received 24 March 2005 / Accepted 26 April 2005}

\abstract{Starspots are usually associated with the action of magnetic fields at the stellar surfaces.
However, recently an inhomogeneous chemical distribution of mercury was found for the mercury-manganese
(HgMn) star \aand\ -- a well-established member of a \textit{non-magnetic} subclass of the chemically peculiar
stars of the upper main sequence. In this study we present first results of the high-resolution survey of
the \hglin\ resonance line in the spectra of rapidly rotating HgMn stars with atmospheric parameters
similar to those of \aand. We use spectrum synthesis modelling and take advantage of the Doppler
resolution of the stellar surfaces to probe horizontal structure of mercury distribution. Clear
signatures of spots are found in the \hglin\ line profiles of HR\,1185 and HR\,8723. Two observations of
the latter star separated by two days give evidence for the line profile variability. We conclude that 
inhomogeneous distribution of Hg is a common phenomenon for the rapidly rotating HgMn stars in the 
13\,000--13\,800~K 
effective temperature range independently of the stellar evolutionary stage. These results establish
existence of a new class of spectrum variable spotted B-type stars. It is suggested that the
observed Hg inhomogeneities arise from dynamical instabilities in the chemical diffusion processes and 
are unrelated to magnetic phenomena.
}

\maketitle

\section{Introduction}
\label{intro}

In several classes of the main sequence stars surface structures are created and supported by
a strong magnetic field. The Sun and late-type active stars show complex surface magnetic topologies
where the field suppresses convective energy transport and creates cool dark spots. These structures
evolve on a relatively short timescale, whereas the magnetic field itself is regenerated by a
contemporary dynamo mechanism and undergoes quasi-periodic evolution during activity cycles lasting from
a few years to decades. In contrast, some of the B, A, and F chemically peculiar stars have strong
global, often dipolar, fields of $\sim$\,1--10~kG strength which do not change on the short timescales and
are believed to be fossil remnants from the epoch of stellar formation. These strong fields introduce
anisotropy in the processes of radiative diffusion, resulting in depletion and accumulation of certain
chemical elements in different depth layers and surface zones (Michaud et al. \cite{MCM81}). Since 
magnetic field geometry and uneven surface chemical distribution are generally non-symmetric with
respect to the rotation axis, many upper main sequence magnetic stars exhibit rotational modulation of
various magnetic field observables, line profiles, and brightness in broad photometric bands -- a
behaviour commonly described by \textit{the oblique rotator model}.

Alongside the magnetic spotted chemically peculiar stars there exists a group of 
mercury-manganese (HgMn) stars for which no solid evidence of a magnetic field is detected (Shorlin
et al. \cite{SWD02} and references therein). These objects are found within the main sequence band in
the \te\ range between 10\,500 and 15\,000~K and have extremely stable photospheres, providing one of
the best astrophysical laboratories for investigation into the radiative diffusion in stars. None of
the variability phenomena associated with the oblique rotator framework has been unambigously
detected nor expected for the HgMn stars until recently. This paradigm had to be abandoned when 
convincing evidence for the profile changes in the resonance \hglin\ line has been revealed by the very
high quality observations of the brightest HgMn star \aand\ collected by Ryabchikova et al.
(\cite{RMA99}) and Adelman et al. (\cite{AGK02}). The authors of the latter study demonstrated that
the most likely interpretation of this variability is an inhomogeneous Hg surface distribution, coupled
with the 2.382~day rotation period of the star. Doppler mapping of the mercury distribution in \aand\ 
(Adelman et al. \cite{AGK02}) showed that this element is depleted close to both rotation poles and is
concentrated in a series of spots located at the stellar equator.

The discovery of Hg spots in the atmosphere of \aand\ has challenged our understanding of the nature
of HgMn stars and the role of magnetic field in creating stellar surface structures. In no classical
magnetic A or B star with inhomogeneous chemical distributions could we find a spotted structure limited
to just one chemical element and showing such a high degree of symmetry with respect to the stellar
rotation axis. Adding to the puzzle, numerous high-precision spectropolarimetric observations of \aand\
(Glagolevskij et al. \cite{GRB85}; Chountonov \cite{C01}; Wade et al. \cite{WAA04}) failed to detect
magnetic field which could be responsible for the observed anomalous behaviour of Hg.

A clue to the nature of the \aand\ phenomenon can be obtained by establishing its extent among
other HgMn stars. This requires collecting high-resolution high $S/N$ observations of Hg lines followed
by a careful time-series analyses of the integral line profile characteristics, such as equivalent
width, line depth, and central wavelength. A different and more sensitive type of observation can be performed
for the rapidly rotating HgMn stars. In these objects the stellar line shapes are dominated by the rotational
Doppler broadening and spectral contribution of different surface zones could be resolved in the
profiles of absorption lines. 
In this situation surface structures manifest themselves through significant deviation of the spectral 
line shapes from a pure rotational profile. 
The \textit{Doppler imaging} method (Vogt \& Penrod \cite{VP83}) exploits
these unique properties of the spectra of rapidly rotating stars and, in principle, is capable
of establishing the presence of stellar surface inhomogeneities from a single
high-quality recording of a spectral line profile.

In the present study we utilize the Doppler imaging principle and search for the signatures of mercury
spots in the atmospheres of rapidly rotating HgMn stars with \te\ similar to that of \aand. Using this
approach we discover two HgMn stars with inhomogeneous surface distribution of Hg and demonstrate line
profile variability for one of them.

\section{Observations and data reduction}
\label{observ}

We recorded high-resolution spectra of rapidly rotating HgMn stars using the cross-dispersed Nasmyth
Echelle Spectrometer (NES, Panchuk et al. \cite{NES}) installed at the 6-m telescope of the Russian
Special Astrophysical Observatory. A sample of 6 bright ($V$\,=\,5.45--6.05) HgMn stars was chosen
according to the spectral classification in the SIMBAD database and literature references. 
Catalogues of the projected rotational velocity measurements (Abt et al. \cite{ALG02}; Strom et al.
\cite{SWD03}) were used to select HgMn stars with \vs\,$\ga$\,50~\kms. According to the literature
references in SIMBAD and information in the relevant catalogues (Pourbaix et al. \cite{PTB04}), none of
the program stars is known to be a spectroscopic binary.

Stellar observations were obtained during the period between July 28 and August 5, 2004. All program
stars were observed once, except HR\,8723 for which two spectra were obtained. A 2K$\times$2K Loral
CCD detector used with the NES spectrograph allowed us to record 22 echelle orders with complete
wavelength coverage of the 3495--4190~\AA\ region and a spectral resolving power of
$\lambda/\Delta\lambda=36\,500$. 

\begin{table}[!t]
\caption{The log of NES observations of 
rapidly rotating HgMn stars. Columns give the HR and HD numbers of the target stars,
heliocentric Julian Date for the middle of exposure, the length of exposure, and
the signal-to-noise ratio achieved for the spectral region around the \hglin\ line.}
\label{tbl1}
\begin{tabular}{ccccc}
\hline
\hline
HR & HD & HJD$-$2\,400\,000 & $T_{\rm exp}$ (min) & $S/N$ \\
\hline
1079 &  21933 & 53223.506 & 50 & 290  \\
1185 &  23950 & 53218.519 & 55 & 370  \\
1445 &  28929 & 53215.532 & 60 & 330  \\
1484 &  29589 & 53219.548 & 40 & 330  \\
1576 &  31373 & 53224.523 & 60 & 290  \\
8723 & 216831 & 53216.520 & 40 & 510  \\
     &        & 53218.432 & 40 & 470  \\
\hline
\end{tabular}
\end{table}

The software package {\sc reduce} (Piskunov \& Valenti \cite{PV02}) was used for the optimal extraction
of the NES spectra. A wavelength scale with an internal accuracy of 50--70~\ms\ was established by means
of a 2-D wavelength calibration procedure which made use of $\approx$\,600 ThAr lines in all echelle
orders. Provisional normalization of the spectra was performed by fitting a smooth function to the
line-free spectral regions. This procedure is not applicable to the echelle orders containing
wide hydrogen Balmer lines. Continuum in these regions was instead adjusted in such a way that the observed
hydrogen line wings would match synthetic spectra computed as described in Sect.~\ref{synthesis}.

For the employed CCD detector and spectral resolution of our observations the instrumental profile
of the spectrograph is somewhat oversampled (4.4 pixels/resolution element). This allowed us to reduce
the pixel sampling of the  spectra by a factor of 2 which increased the $S/N$ ratio without compromising
resolution. The signal-to-noise ratio of the resampled spectra lies in the range of 290--510 for the
spectral region containing the \hglin\ line.

The summary of our spectroscopic observations of HgMn stars is given in Table~\ref{tbl1}.

\section{Spectrum synthesis calculations}
\label{synthesis}

To probe the structure of mercury distribution on the surfaces of HgMn stars we have to develop
an objective method for characterizing deviation of the observed profile of the resonance \hg\ line
at $\lambda$~3984~\AA\ from the basic rotational profile. We choose to compare the observed spectra
with the theoretical spectrum synthesis computation performed under assumption of a homogenous Hg
abundance distribution. 

The colours in the Str\"omgren photometric system were extracted from the
$uvby\beta$  catalogue of Hauck \& Mermilliod (\cite{HM98}) and were used to determine atmospheric
parameters,  \te\ and \lgg, with the help of two widely used photometric calibrations (Moon \&
Dworetsky \cite{MD85}; Napiwotzki et al. \cite{NSW93}) implemented in the \mbox{\sc templogg} code
(Rogers \cite{R95}). For all program stars we found consistent atmospheric parameters and adopted
effective temperature and surface gravity in between the values given by the two calibrations (see
Table~\ref{tbl2}). The errors in \te\ and \lgg\ are estimated to be 200~K and 0.1~dex respectively.
Model atmospheres of the program stars were generated with the {\sc atlas9} code of Kurucz (\cite{K93})
assuming zero microturbulence and solar chemical composition. 

\begin{table*}
\caption{Atmospheric parameters and other characteristics of the
rapidly rotating HgMn stars included in our study. Columns give the HR number of the program stars,
effective temperature and surface gravity obtained from the Str\"omgren photometric
colours using calibrations of Moon \& Dworetsky (\cite{MD85}) and Napiwotzki et al. 
(\cite{NSW93}), \te\ and \lgg\ adopted in the present paper, Hipparcos parallax, and
projected rotational velocity. The last three columns report inferred abundances
of Si, Y, and Hg. The latter parameter corresponds to the optimum fit of the \hg\ line wings.}
\label{tbl2}
\begin{tabular}{ccccccccrccc}
\hline
\hline
HR & \multicolumn{2}{c}{Moon \& Dworetsky (\cite{MD85})} & 
     \multicolumn{2}{c}{Napiwotzki et al. (\cite{NSW93})} & 
     \multicolumn{2}{c}{Adopted} & \vs\ & $\pi$~~~~~~~ & \multicolumn{3}{c}{$\log N/N_{\rm tot}$} \\
   & \te\ (K) & \lgg\ & \te\ (K) & \lgg\ & \te\ (K) & \lgg\ & (\kms) & (mas)~~~~ & Si & Y & Hg \\
\hline
1079 & 11948 & 4.13 & 12040 & 4.07 & 12000 & 4.1 & 87.6 &  9.18$\pm$0.87 & $-5.00$ & $-6.60$ & $-6.00$ \\
1185 & 12892 & 4.06 & 13110 & 3.97 & 13000 & 4.0 & 67.0 & 10.14$\pm$0.90 & $-4.80$ & $-7.50$ & $-5.45$ \\
1445 & 12885 & 4.05 & 12990 & 3.97 & 12900 & 4.0 & 56.4 &  7.00$\pm$0.86 & $-4.40$ & $-7.80$ & $-6.15$ \\
1484 & 14473 & 4.29 & 14763 & 4.16 & 14600 & 4.2 & 63.2 &  9.46$\pm$0.78 & $-4.70$ & $-7.50$ & $-6.80$ \\
1576 & 13884 & 4.14 & 14013 & 4.03 & 13900 & 4.1 & 74.6 &  7.71$\pm$0.87 & $-4.50$ & $-7.50$ & $-6.20$ \\
8723 & 13057 & 3.65 & 13058 & 3.56 & 13100 & 3.6 & 67.1 &  3.90$\pm$0.70 & $-4.65$ & $-7.30$ & $-6.40$ \\
\hline							   
\end{tabular}
\end{table*}

The spectrum synthesis calculations in this paper were carried out with the help of the {\sc synth3}
code developed by O. Kochukhov. This program is based on the {\sc synth} code (Piskunov \cite{P92})
and {\sc sme} package (Valenti \& Piskunov \cite{VP96}). {\sc synth3} uses a non-magnetic version of
the quadratic DELO radiative transfer algorithm (Socas-Navarro et al. \cite{STR00}) to compute
theoretical intensity spectra on the adaptive wavelength grid for a range of angles between the line
of sight and normal to the stellar surface. The disk-integrated flux spectrum is calculated for a
given \vs\ by summing the Doppler-broadened contributions of different surface annular zones. This
spectrum synthesis approach allows us to reach a higher precision in modelling the spectra of rapidly
rotating stars compared to the convolution of the flux spectra with a Doppler broadening function
(Gray \cite{G92}) assuming a linear limb-darkening law.

Numerical procedure for the calculation of the opacity due to the hydrogen Balmer lines implemented
in {\sc synth3} includes recent developments in the Balmer line broadening theory (Stehl\'e
\cite{S94}; Barklem et al. \cite{BPO00}).

Atomic line list used in our study of the HgMn stars was obtained from the VALD database (Kupka et al.
\cite{KPR99}). In addition, the line list was complemented by 15 isotope and hyperfine components of
the \hglin\ line based on the data given in Table~3 of Woolf \& Lambert (\cite{WL99}). The oscillator
strengths of the \hg\ line components were recalculated for the terrestrial isotope mixture of Hg (see
Table~1 in Proffitt et al. \cite{PBL99}).

\begin{figure}[!th]
\figps{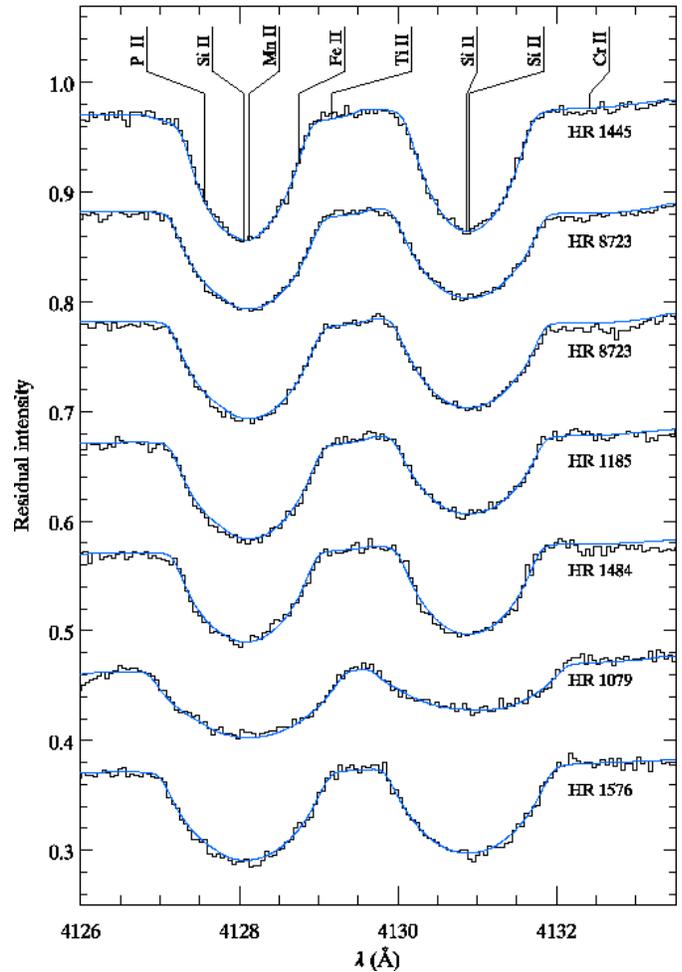}
\caption{The spectra of rapidly rotating HgMn stars in the vicinity of the \ion{Si}{ii} 4128.05
and 4130.89~\AA\ lines. Observations are shown with histograms, whereas the solid curves correspond to
the best-fit synthetic spectra. Stellar spectra are displaced in the vertical direction
for display purpose. Two observations of HR\,8723 are compared with the same theoretical calculation.}
\label{fig1}
\end{figure}

\subsection{\ion{Si}{ii} lines}
\label{silicon}

Since all HgMn stars investigated here are characterized by substantial rotation, \vs\ is the most 
important parameter determining the spectral line shapes. Accurate measurements of the projected
rotational velocity were obtained by fitting a short spectral segment containing strong \ion{Si}{ii}
lines at $\lambda$ 4128.05, 4130.87, 4130.89~\AA. The stellar radial velocity, \vs, and the silicon
abundance were adjusted simultaneously. Occasionally, concentrations of Mn, Ti, and Cr were also
modified to obtain a better description of the weak features of these elements blending the
\ion{Si}{ii} lines. Resulting \vs\ parameter and Si abundance are reported in Table~\ref{tbl2}. These values are
uncertain to approximately $\pm1$~\kms\ and $\pm$0.05~dex respectively.

The fit of the high $S/N$ spectra of HgMn stars in the 4126--4134~\AA\ region is illustrated in
Fig.~\ref{fig1}. For all 7 observations of 6 stars we are able to obtain a very good description of the
\ion{Si}{ii} line shapes. No distortion of the \ion{Si}{ii} lines indicative of the surface structures
or spectroscopic binarity is detected.

\begin{figure}[!t]
\figps{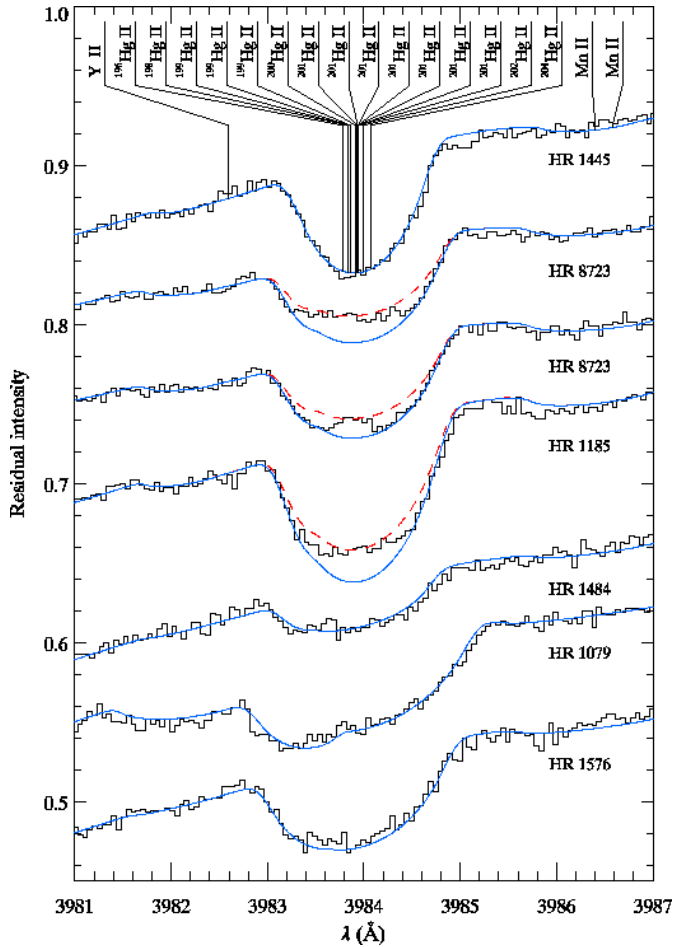}
\caption{The resonance \hglin\ in the spectra of rapidly rotating HgMn stars. Observations are
shown with histogram, whereas lines correspond to the best-fit theoretical spectra. The solid curve
illustrates the fit of the \hg\ line wings. For the spectrum of HR\,1185 and two observations of
HR\,8723 showing anomalous line profile shapes additional theoretical spectrum (dashed line) shows
approximate fit of the line cores. The spectra of the HgMn stars are displaced in the vertical direction
for display purpose.}
\label{fig2}
\end{figure}

\subsection{\ion{Hg}{ii} line}
\label{mercury}

The resonance line of \hg\ at $\lambda$~3984~\AA\ was analysed using the \vs\ value determined from
fitting the \ion{Si}{ii} lines. To achieve a good description of this spectral region abundances of Hg, Y,
and Mn were adjusted simultaneously. Absorption due to the red wing of the hydrogen H$\varepsilon$ line
was taken into account. Resulting best-fit synthetic spectra are illustrated in
Fig.~\ref{fig2}. Inferred abundances of Hg and Y are given in Table~\ref{tbl2}. The respective
internal error bars are approximately 0.1 and 0.3~dex.
For HR\,1445, HR\,1484, HR\,1079, and HR\,1576 the observed spectra can be reproduced
within the error bars with a unique Hg abundance. We note that the double structure visible in the \hg\
line of HR\,1079 is due to the blending with the unusually strong \ion{Y}{ii} 3982.59~\AA\ line and is
successfully modelled by our theoretical calculation. At the same time, we find that the \hg\ line in the
spectra of HR\,1185 and HR\,8723 could not be fitted assuming a homogeneous surface distribution of
mercury. When Hg abundance is chosen to describe the outer line wings, the inner part of the profile
is too deep compared to observations. Decreasing abundance to fit intensity in the line core does not produce
acceptable results either. It is clear that the overall shape of the \hglin\ line deviates significantly
from the Doppler-broadened profile expected for these two HgMn stars.

It should be emphasized that the wavelength separation of the isotope components of the \hg\ line is
considerably smaller than Doppler broadening for \vs\,$=$\,50--70~\kms. Hence, an unusual isotope
mixture could not be responsible for the observed line shapes. Our spectrum synthesis calculations
demonstrate that, even in the extreme case when the \hg\ line contains only the $^{196}$Hg and $^{204}$Hg
isotope components, a flat-bottomed spectral line shape is only seen for \vs\ not exceeding
$\approx$\,15~\kms. 

The shallow \hg\ line profiles of HR\,1185 and HR\,8723 bear strong resemblance to the appearance of
this line in some of the rotation phases of \aand\ (see Fig.~5 in Adelman et al. \cite{AGK02}).
Consequently, the line profile anomalies found for HR\,1185 and HR\,8723 are interpreted as a signature
of an inhomogeneous surface distribution of Hg.

Two observations of HR\,8723 separated by 2 days are shown in Fig.~\ref{fig2}. The \hg\ line profile has
clearly changed, whereas the \ion{Si}{ii} lines (Fig.~\ref{fig1}) and all other spectral features are
identical within the noise level in the two spectra. This behaviour has a striking similarity to the
spectrum variability pattern of \aand\ and strengthens the conclusion that the \hg\ line anomalies seen
in HR\,8723 arise from a spotted distribution of Hg and that the corresponding profile variability occurs with the
rotation period of the star.

\section{Discussion}
\label{discussion}

Inspired by the recent discovery and mapping of Hg spots in the atmosphere of \aand\ (Adelman et al.
\cite{AGK02}), we carried out a high $S/N$ survey of the \hglin\ line in a carefully chosen group of 
rapidly rotating HgMn stars. For such objects Doppler broadening of the stellar spectra allowed us to
obtain partial resolution of the stellar surface and probe possible Hg inhomogeneities.  Our observations
demonstrate that a spotted Hg distribution is not limited to \aand\ and exists in other HgMn stars.  With
the help of detailed spectrum synthesis calculations we revealed signatures of an inhomogeneous mercury
distribution in two out of six studied stars. Furthermore, for the HgMn star HR\,8723 we discover profile
variability of the \hg\ line. These observations establish existence of a new class of spectrum
variable, spotted, apparently non-magnetic, HgMn stars. The three currently known members include \aand,
HR\,1185, and HR\,8723.

\begin{figure}[!t]
\figps{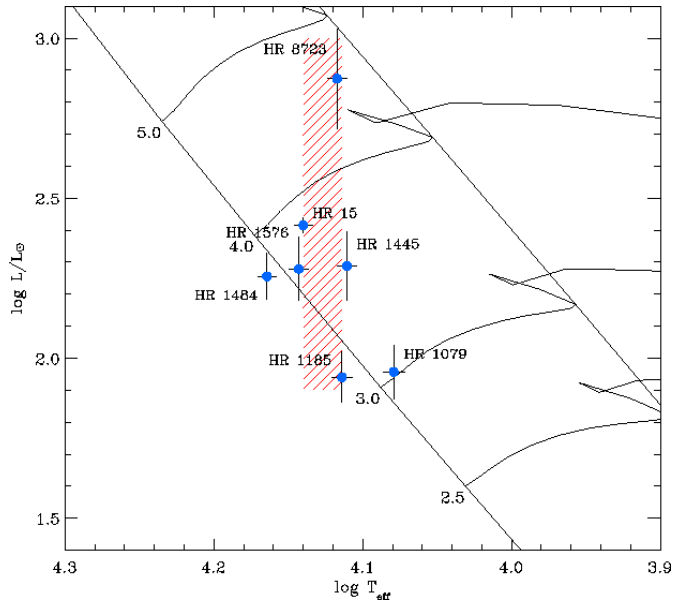}
\caption{Effective temperatures and luminosities of rapidly rotating HgMn stars (symbols) are 
compared with the theoretical evolutionary tracks of Schaller et al. (\cite{SSM92}). The shaded area
highlights the 13\,000--13\,800~K \te\ range where HgMn stars with an inhomogeneous mercury surface
distribution were found so far.}
\label{fig3}
\end{figure}

Precise Hipparcos parallaxes (ESA \cite{ESA97}) reported for the program stars in
Table~\ref{tbl2} enable us to establish evolutionary status of the HgMn stars investigated here. In
Fig.~\ref{fig3} we compare effective temperatures and luminosities of the HgMn stars studied by us and \aand\
(=HR\,15, \te\,$=$\,13\,800~K, Ryabchikova et al. \cite{RMA99}) with the theoretical stellar evolutionary
tracks of Schaller et al. (\cite{SSM92}). Most of the stars are located close to the zero-age main sequence
line and have masses between 3.0 and 4.0~\msun. HR\,8723 is apparently a more massive and evolved HgMn star
with $\cal{M}$\,$\approx$\,4.5~\msun. There seems to be no relation between the stellar age and
presence of mercury spots. At the same time, a striking result is that the spotted HgMn stars occupy
\textit{a relatively narrow range of \te\ between 13\,000 and 13\,800~K}. Further observations are required
to confirm this intriguing behaviour, but taken at face value the outcome of our study indicates that 
Hg spots can only form in the atmospheres of rapidly rotating HgMn stars lying in this limited \te\
interval. 

The origin of Hg inhomogeneities in the atmospheres of HgMn stars remains a mystery. The standard link
between stellar surface structures and magnetic field is doubtful in this case  because the HgMn stars
as a class are known to be non-magnetic (Shorlin et al. \cite{SWD02}). Numerous attempts to detect magnetic
field in \aand\ itself (Glagolevskij et al. \cite{GRB85}; Chountounov \cite{C01}; Wade et al. \cite{WAA04})
steadily decrease an upper limit on the strength of the possible global field, culminating in the recent
unprecedented longitudinal field measurement consistent with zero magnetic field, 
with an error bar less than 10~G, obtained with the new ESPaDOnS spectropolarimeter at
CFHT by G.~Wade and collaborators (private communication). This places a stringent upper limit of a few tens G
on the possible large-scale field, which is well below the equipartition limit for \aand. Furthermore, the
high-resolution spectropolarimetric observations by Wade et al. (\cite{WAA04}) also place severe constraints on
the possible presence of higher-order multipole and tangled fields -- a type of magnetic topology which has
been successfully detected using identical observational methods in the Doppler-broadened Stokes $V$ profiles
of active stars (Donati et al. \cite{DSC97}). Thus, if we assume that the spotted HgMn stars are magnetic at
some low level, their field topology as well as the origin of magnetism have to be very unusual and
qualitatively different from both global fossil fields of the early-type chemically peculiar stars and
complex dynamo-generated magnetic structures found in the active late-type stars. At the same time, these
hypothetical magnetic fields of HgMn stars have to possess some degree of global coherency in order to be
related to the observed large-scale Hg structures. At the current level of understanding it appears exceedingly
difficult to reconcile these contradictory requirements on the field topology and, therefore, we consider a
magnetic explanation of the Hg inhomogeneities to be very unlikely.

An alternative interpretation may be related to the possible dynamical instability of the diffusion processes
in rapidly rotating HgMn stars. It is not inconceivable that in the narrow range of stellar parameters
where Hg spots are found accumulation of this element in the line forming region becomes extremely
sensitive to external perturbations. In this situation a few \% reduction of the effective gravitational 
acceleration at the equators of the HgMn stars with substantial rotation is sufficient to produce the observed
pole-to-equator gradient in the Hg concentration. In addition to this axisymmetric effect, a range of
hydrodynamical instabilities, possibly related to the rotationally induced mixing of the subphotospheric
layers, may inhibit or amplify a buildup of the Hg overabundance in different surface zones, creating a
patchy time-dependent horizontal distribution of this chemical element. Detailed theoretical investigations
into the Hg diffusion are needed to verify this hypothesis and establish timescales of the possible evolution
of Hg distribution. A long-term observational campaign aimed at detecting such changes in the surface Hg
maps also appears to be extremely promising by itself and as a test of predictions of the radiative diffusion 
theory.

\begin{acknowledgements}

We acknowledge resources provided by the Vienna Atomic Line
Database (VALD) and the SIMBAD database, operated at CDS, Strasbourg, France.

We also thank the referee, Dr. G. Wade, whose constructive remarks
contributed to the improvement of this paper.

\end{acknowledgements}

\end{document}